\documentclass[12pt]{iopart}
\usepackage{times}
\usepackage{mathptmx}
\usepackage{graphicx}
\usepackage[dvips,colorlinks]{hyperref}

\begin{document}

\centerline { \bf FRENET-SERRET VACUUM RADIATION, DETECTION PROPOSALS} 
\centerline{\bf AND RELATED TOPICS}

\bigskip 

\centerline {Haret C. Rosu$^1$\footnote{hcr@ipicyt.edu.mx}} 

\bigskip

\centerline{$^1$ Dept. of Applied Mathematics} 
\centerline{Potosinian Institute of Scientific and Technological Research}
\centerline{Apdo Postal 3-74 Tangamanga, San Luis Potos\'{\i}, MEXICO} 


\bigskip

\centerline {Dated: January 2003} 


\bigskip

\noindent
{
{\bf Abstract} 

\noindent
The paradigmatic Unruh radiation is an ideal and simple case of stationary vacuum radiation patterns related to worldlines defined
as Frenet-Serret curves. We review the corresponding body of literature as well as the experimental proposals that have been suggested
to detect these types of quantum field radiation patterns. Finally, we comment on a few other topics related to the Unruh effect.\\}   
 
\bigskip
\bigskip
\bigskip
\bigskip

Part 1. Frenet-Serret Vacuum Radiation: pp. 2-11

\bigskip
\bigskip

Part 2. Detection Proposals: pp. 12-22

\bigskip
\bigskip

Part 3. Related Topics: pp. 23-26

\bigskip
\bigskip 

References: pp. 27-28

\bigskip
\bigskip
\bigskip
{\em extended version of invited talk at QABP3, Hiroshima, Japan, January 7-11, 2003}

\noindent

\newpage

\section{\bf FRENET-SERRET VACUUM RADIATION}


\bigskip

\noindent
J.F. Frenet (1816-1900) wrote a doctoral thesis in 1847. Part of it deals with the theory of space curves and contains six formulas of the nine three-dimensional
Frenet-Serret formulas. He published his results in {\em Journal de mathematiques pures et appliques} in 1852. J.A. Serret (1819-1885) gave the set of all nine
formulas in three dimensions.

\bigskip
\noindent
The standard FS formulas are the following
\begin{eqnarray*} 
\dot{\vec{T}} & = &\kappa \cdot \vec{N}\\
\dot{\vec{N}} & = &\tau \cdot \vec{B}-\kappa \cdot \vec{T}\\
\dot{\vec{B}} & = &-\tau \cdot \vec{N}~.\\
\end{eqnarray*}

\noindent
These three formulas give the derivatives of the unit tangent, normal, and binormal vectors, respectively, as vectorial combinations of the moving basis in which the two Frenet-Serret invariants
of the normal three-dimensional space, curvature and torsion, respectively, enter as scalar coefficients.
They can be written in matrix form as well
\begin{equation*} \label{matrix}
\left( \begin{array}{ccc}
\dot{\vec{T}}\\
\dot{\vec{N}}\\
\dot{\vec{B}}
\end{array} \right )
=
\left(\begin{array}{ccc}
 0 & \kappa & 0 \\
-\kappa & 0 & \tau\\
0 & -\tau & 0\end{array} \right )\left( \begin{array}{ccc}
\vec{T}\\
\vec{N}\\
\vec{B}
\end{array}\right)
\end{equation*}

\bigskip
\noindent
The plane formed by the span of $\vec{T}$ and $\vec{N}$ is called the osculating plane. The span of $\vec{N}$ and $\vec{B}$ is the normal plane and 
the span of $\vec{B}$ and $\vec{T}$ is the transverse plane to the curve at the point where the Frenet-Serret triad is considered.

\bigskip
\noindent
For the goals of the first part of this review, the interested reader may consult several papers of more recent times. These are seminal papers that
apply the Frenet-Serret formalism to relativistic world lines and quantum field theory. 
First, as quoted by J.L. Synge, Jasper \cite{jas47} studied in 1947 helices in flat space of $n$ dimensions, but only for a positive-definite metric.
Next, Synge wrote an important paper on timelike helices in flat space-time and classified them in six types according to the relative magnitudes of 
the three Frenet-Serret invariants \cite{syng67}. He also proved for the first time that the world line of a charged particle moving in a constant electromagnetic field
is a helix. A few years later, in the seminal paper of Honig, Schucking and Vishveshwara \cite{hsv74} a connection was established between the Lorentz
invariants of the elctromagnetic field and the Frenet-Serret invariants. The 1981 Letaw's paper \cite{l81} is in fact the first work in which  the FS invariants 
and quantum field concepts are merged together (see below). Interesting generalizations to more dimensions, black hole environments, and gyroscopic 
precession have been investigated by Iyer and Vishveshwara \cite{iv} (see also \cite{bjm}).

\bigskip
\noindent
We also notice that at  the strictly classical level, the Frenet-Serret coordinate system is quite used in accelerator physics to express
Hamiltonians and/or fields \cite{chao}. 

\bigskip


\subsection{\bf  Stationary world lines and the vacuum excitation of noninertial detectors}

\bigskip

\noindent
In 1981, Letaw studied the  
stationary world lines, on which quantized field detectors in a scalar vacuum have time-independent excitation
spectra. They are characterized by the requirement that the geodetic interval between two points depends only
on the proper time interval. He used a generalization of the Frenet equations to Minkowski space and found, not surprisingly, that the
curvature invariants are the proper acceleration and angular velocity of the world line. Solving the generalized Frenet equations for the simple case of
constant invariants leads to  several classes of stationary world lines. Letaw gave a classification into six types reviewed here. He also demonstrated the
equivalence of the timelike Killing vector field orbits and the stationary world lines. 
Last but not least, Letaw did some calculations of the vacuum excitation spectra of detectors on the sample of six families of stationary world lines.
Letaw's work is a generalization of Unruh's famous result concerning the excitation of a scalar-particle detector moving with constant linear acceleration in the vacuum of flat spacetime. 
According to Unruh, the detector behaves as if in contact with a bath of scalar particles with energies in a Planck spectrum of temperature: acceleration/2$\pi$. 
Because of the connection with the Hawking radiation and its paradigmatic nature, the Unruh effect attracted the attention of many theoretical physicists.
On the other hand, the present author considers that Letaw's results place the Unruh effect in a different and more general perspective that is elaborated herein.
We hope that a more effective attitude is put forward in the present review that could be more rewarding in the years to come and may help to strengthen the links
between laboratory physics and astrophysics.


\bigskip
\noindent
According to DeWitt, 
the probability for a detector moving along a world line $x^{\mu}(\tau)$ to be found in an excited state of energy $E$ at $\tau=\tau _0$ is given in terms 
of the autocorrelation function of the field (Wightman function) 
\begin{equation}\label{l81-01}
P(E)=D(E)\int _{-\infty}^{\tau _0}d\tau \int _{-\infty}^{\tau _0}d\tau ^{'}e^{-iE(\tau -\tau ^{'})}\,\langle 0|\phi(x(\tau))\phi(x(\tau ^{'}))|0\rangle~,
\end{equation}
where $D(E)$ is a function characterizing the sensitivity of the detector. The Wightman function for a scalar field is proportional to the inverse of the geodetic 
interval $\Delta (\tau , \tau ^{'})$
\begin{equation} \label{l81-02}
 \langle 0|\phi(x(\tau))\phi(x(\tau ^{'}))|0\rangle =
[2\pi ^2\Delta (\tau , \tau ^{'})]^{-1}~,
\end{equation}
where 
\begin{equation}\label{l81-01b}
\Delta (\tau , \tau ^{'})=[x_{\mu}(\tau)-x_{\mu}(\tau ^{'})][x^{\mu}(\tau)-x^{\mu}(\tau ^{'})]~.
\end{equation}
The rate of excitation to the state with energy $E$ is the Fourier cosine transform
\begin{equation}\label{l81-03}
\frac{dP(E)}{d\tau _0}=2D(E)\int _{-\infty}^{0}ds\langle 0|\phi(x(\tau _0))\phi(x(\tau _0+s))|0\rangle\cos(Es)~.
\end{equation}
where $s=\tau -\tau ^{'}$ is the proper time interval. The rate of excitation (the response) is directly related to the energy spectrum of the detected scalar
``particles"
\begin{equation} \label{l81-04}
S(E, \tau)=2\pi \rho (E) \int _{-\infty}^{0}ds\langle 0|\phi(x(\tau))\phi(x(\tau +s))|0\rangle\cos(Es)~.
\end{equation}
If the Wightman function is time independent the detected spectra are stationary. This is equivalent to the following property of the geodetic interval
\begin{equation}\label{l81-05}
\Delta (\tau, \tau +s)=\Delta (0,s)~.
\end{equation}

\bigskip

\newpage

{\em Curvature invariants and the Frenet equations}

\bigskip
\noindent
An arbitrary timelike world line in flat space is generally described by four functions, $x^{\mu}(s)$,
specifying the coordinates of each point $s$ on the curve. This parameter may be taken to be the arc length or proper time on the world line.
The parametric representation is unsatisfactory in two respects: 

\bigskip

\noindent
(1) {\em A world line is a geometric object and should not require a coordinate-dependent entity for its definition}.

\bigskip
\noindent
(2) {\em There is an inherent redundancy in the parametric representation since three functions suffice to determine the world line}.

\bigskip

\noindent
The Frenet-Serret (curvature) invariants, on the other hand, provide an intrinsic definition of the world line not subject to these criticisms.

\bigskip
\noindent
The starting point is the construction of an orthonormal tetrad $V_{a}^{\mu}(s)$ at every point on the world line $x^{\mu}(s)$. The Latin index everywhere is a tetrad
index. The tetrad is formed from the derivatives of $x^{\mu}(s)$ with respect to proper time (represented by one or more dots). It is assumed that the 
first four derivatives are linearly independent, the results being practically unchanged when they are not. Members of the tetrad must satisfy the 
orthonormality condition
\begin{equation} \label{l81-1}
V_{a\mu}V_{b}^{\mu}=\eta _{ab}~,
\end{equation}
where the metric has diagonal components (1,-1,-1,-1) only.

By Gram-Schmidt orthogonalization of the derivatives working upwords from the first, the following expressions for the tetrad are found:
\begin{equation} \label{l81-2}
V_{0}^{\mu}=\dot x^{\mu}~,
\end{equation}
\begin{equation} \label{l81-3}
V_{1}^{\mu}=\frac{{\ddot x}^{\alpha}}{(-\ddot x_{\alpha}\ddot x^{\alpha})^{1/2}}~,
\end{equation}
\begin{equation} \label{l81-4}
V_{2}^{\mu}=\frac{(\ddot x_{\gamma}\ddot x^{\gamma})\stackrel{...}x^{\mu}- (\ddot x_{\gamma}\stackrel{...}x^{\gamma})\ddot x^{\mu} +(\ddot x_{\gamma}\ddot x^{\gamma})^2\dot x ^{\mu}}
{[(\ddot x_{\alpha}\ddot x^{\alpha})^4+(\ddot x_{\alpha}\ddot x^{\alpha})(\ddot x_{\beta}\stackrel{...}x^{\beta})^2-
(\ddot x_{\alpha}\ddot x^{\alpha})^2(\stackrel{...}x^{\beta}\stackrel{...}x^{\beta})]^{1/2}}~,
\end{equation}
\begin{equation} \label{l81-5}
V_{3}^{\mu}=\frac{1}{\sqrt{6}}\epsilon ^{\alpha \beta \gamma \mu}V_{0\alpha}V_{1\beta}V_{2\gamma}~,
\end{equation}
Overall signs on these vectors are fixed by the orientation of the tetrad.

The tetrad $V_{\sigma}^{\mu}$ is a basis for the vector space at a point on the world line. Derivatives of the basis vectors may therefore be expanded in terms of 
them:
\begin{equation} \label{l81-6}
\dot V _{a}^{\mu}=K_{a}^{b}V_{b}^{\mu}~.
\end{equation}
These are the generalized Frenet equations. $K_{ab}$ is a coordinate-independent matrix whose structure must be determined.

Differentiation of the orthonormality condition (\ref{l81-1}) yields
\begin{equation} \label{l81-7}
\dot V_{a\mu}V_{b}^{\mu}+V_{a\mu}\dot V_{b}^{\mu}=0~,
\end{equation}
and, in view of (\ref{l81-6}),
\begin{equation} \label{l81-8}
K_{ab}=-K_{ba}~.
\end{equation}
A basis vector $V_{a}^{\mu}$ is defined in terms of the first $a+1$ derivatives of $x^{\mu}$; therefore, $\dot V_{a}^{\mu}$ will be 
a linear combination of the first $a+2$ derivatives. These $a+2$ derivatives are dependent only on the basis vectors $V_{b}^{\mu}$
where $b\leq a+1$. This and (\ref{l81-8}) limit the matrix to the form
\begin{equation}\label{l81-9}
K_{ab}=
\left(\begin{array}{cccc}
 0 & -\kappa (s)& 0 & 0\\
\kappa (s)& 0 & -\tau _1 (s)& 0\\
0 & \tau _1 & 0 & -\tau _2\\
0 & 0 & \tau _ 2 & 0 \end{array} \right )
\end{equation}
The three functions of proper time are the invariants
\begin{equation} \label{l81-10}
\kappa=V_{0\mu}\dot V_{1}^{\mu}=-\dot V_{0\mu}V_{1}^{\mu}~,
\end{equation}
\begin{equation} \label{l81-11}
\tau _1=V_{1\mu}\dot V_{2}^{\mu}=-\dot V_{1\mu}V_{2}^{\mu}~,
\end{equation}
\begin{equation} \label{l81-12}
\tau _2=V_{2\mu}\dot V_{3}^{\mu}=-\dot V_{2\mu}V_{3}^{\mu}~.
\end{equation}
They are, respectively, the curvature, the first torsion, and the second torsion (hypertorsion) of the world line. Sign choices are made for reasons brought out 
below.

To explore the physical significance of the invariants we examine the infinitesimal Lorentz transformations of the tetrad at a point on the world line. The transformations leave 
the metric invariant
\begin{equation} \label{l81-13}
\eta _{ab}=L_{a}^{c}L_{b}^{d}\eta _{cd}~.
\end{equation}
An infinitesimal transformations may be written 
\begin{equation} \label{l81-14}
L_{a}^{c}=\delta _{a}^{c}+d\epsilon _{a}^{c}~,
\end{equation}
where the elements of $d\epsilon _{a}^{c}$ are small and must satisfy
\begin{equation} \label{l81-15}
d\epsilon _{ab}=-d\epsilon _{ba}~,
\end{equation}
The transformations are taken to be active; that is, the transformed tetrad moves $+v$ and is rotated $\theta$ relative to the untransformed tetrad.
Thus the infinitesimal generator is
\begin{equation} \label{l81-16}
d\epsilon _{ab}=
\left(\begin{array}{cccc}
 0 & -dv _1& -dv_2& -dv _3\\
dv_1& 0 & -d\theta _{12} & d\theta _{31}\\
d v_2 & d\theta _{12} & 0 & -d\theta _{23}\\
dv_3 &-d\theta _{31} & d\theta _{23} & 0 \end{array} \right )
\end{equation}
The change in the tetrad resulting from this transformation is
\begin{equation}\label{l81-17}
\dot V_{a}^{\mu}=(d\epsilon _{a}^{b}/ds)V_{b}^{\mu}~.
\end{equation}
Equations (\ref{l81-17}) are identical to the Frenet equations (\ref{l81-6}); therefore, the physical content of the curvature invariants is found by comparison of 
(\ref{l81-9}) and (\ref{l81-16}). 

\bigskip
{\em Physical interpretation of the FS invariants}

\bigskip

\noindent
{\em  1. $\kappa$ is the proper acceleration of the world line which is always parallel to $V_{1}^{\mu}$.} 

\bigskip
\noindent
{\em 2. $\tau _1$ and $\tau _2$ are the components of proper angular velocity of the world line in the planes spanned by $V_{1}^{\mu}$ and $V_{2}^{\mu}$, and $V_{2}^{\mu}$ and $V_{3}^{\mu}$, respectively. The total proper angular velocity is the vector sum of these two invariants}.

\bigskip
\newpage

{\em Stationary Motions}

\bigskip

\noindent
The simplest worldlines are those whose curvature invariants are constant. They are called {\em stationary} because their geometric properties are independent of 
proper time. 
One also finds that only observers on these world lines may establish a coordinate system in which they are at rest and the metric is stationary. 
Clearly, the geodetic interval between two points on a stationary world line can depend only on the proper time interval, therefore they are the world lines on 
which a detector's excitation is time independent.

\bigskip

\noindent
The Frenet equation (\ref{l81-6}) may be reduced to a fourth-order linear equation in $V_{0}^{\mu}$ when the curvature invariants are constant
\begin{equation} \label{l81-18}
\stackrel{....}V_{0}^{\mu}-2a\ddot V_{0}^{\mu} -b^2 V_{0}^{\mu}=0~,
\end{equation}
where 
$$
a=\frac{1}{2}\left(\kappa ^2-\tau _1^2-\tau _2^2\right), \quad b=|\kappa \tau _2|~.
$$
The other basis vectors are determined from $V_{0}^{\mu}$ by the equations
\begin{equation} \label{l81-19}
V_1^{\mu}=\dot V_{0}^{\mu}/\kappa
\end{equation}
\begin{equation} \label{l81-20}
V_2^{\mu}=(\ddot V_0^{\mu}-\kappa ^2 V_0^{\mu})/\kappa \tau _1
\end{equation}
\begin{equation} \label{l81-21}
V_3^{\mu}=[\stackrel{...}V_0^{\mu}-(\kappa ^2-\tau _1^2)\dot V_0^{\mu}]/\kappa \tau _1 \tau _2~.
\end{equation}

\bigskip

\noindent
Equation (\ref{l81-18}) is homogeneous with constant coefficients. The four roots of the characteristic equation are $\pm R_{+}$ and $\pm i R_{-}$,
where
\begin{equation} \label{l81-22}
R_{\pm}=[(a^2+b^2)^{1/2}\pm a]^{1/2}~.
\end{equation}
The solution is of the form
\begin{equation} \label{l81-23}
V_{0}^{\mu}=A^{\mu}{\rm cosh}(R_{+}s)+B^{\mu}{\rm sinh}(R_{+}s)+C^{\mu}\cos (R_{-}s)+D^{\mu}\sin (R_{-}s)~.
\end{equation}
Using (\ref{l81-19})-(\ref{l81-21}) and (\ref{l81-23}) at $s=0$ and the initial conditions $(V_{a}^{\mu})_{s=0}=\delta _{a}^{\mu}$ one can get the following 
expressions for the coefficients
\begin{eqnarray}
A^{\mu}=R^{-2}(R_{-}^{2}+\kappa ^2, 0, \kappa \tau _1, 0)~,\\
B^{\mu}=R^{-2}(0,\kappa(R_{-}^{2}+\kappa ^2 -\tau ^2)/R_{+},0,\kappa \tau _1 \tau _2/R_{+})~,\\
C^{\mu}=R^{-2}(R_{+}^{2}-\kappa ^2, 0, -\kappa \tau _1, 0)~,\\
D^{\mu}=R^{-2}(0,\kappa(R_{+}^{2}-\kappa ^2 +\tau ^2)/R_{-},0,-\kappa \tau _1 \tau _2/R_{-})~,\\
\end{eqnarray}
with $R^2=R_{+}^{2}+R_{-}^{2}$. From these results on the FS tetrad one gets easily the six classes of stationary world lines and the corresponding 
excitation spectra.

\bigskip
\newpage

{\em The Six Stationary Scalar Frenet-Serret Radiation Spectra}

\bigskip
\noindent
\underline{{\bf 1}. $\kappa =\tau _1=\tau _2=0$} $\qquad$  (inertial, uncurved worldlines)

\bigskip
\noindent
The excitation spectrum is a trivial cubic spectrum 

\bigskip
\begin{equation} 
S_{\rm recta}(E)=\frac{E^3}{4\pi ^2}~, 
\end{equation} 
i.e., as given by a vacuum of zero point energy per mode $E/2$ 
and density of states $E^2/2\pi ^2$.

\bigskip 
\noindent 
\underline{{\bf 2}. $\kappa \neq 0$, $\tau _1=\tau _2=0$} $\qquad$ (hyperbolic worldlines)

\bigskip
\noindent
The excitation spectrum is Planckian allowing the 
interpretation of $\kappa/2\pi$ as `thermodynamic' temperature. In the 
dimensionless variable $\epsilon _{\kappa}=E/\kappa$ the vacuum spectrum reads 

\bigskip
\begin{equation} 
S_{\rm hyp}(\epsilon _{\kappa}) 
=\frac{\epsilon _{\kappa}^{3}}{2\pi ^2(e^{2\pi\epsilon _{\kappa}}-1)}~. 
\end{equation} 

\bigskip 
\noindent 
\underline{{\bf 3}. $|\kappa|<|\tau _1|$, $\tau _2=0$, $\rho ^2=\tau _1 ^2-\kappa ^2$} $\qquad$ (helical worldlines) 

\bigskip
\noindent
The excitation spectrum is an analytic function corresponding to the case 4 below only in the limit $\kappa\gg \rho$ 
\begin{equation} 
S_{\rm hel}(\epsilon _{\rho})\stackrel{\kappa/\rho\rightarrow \infty} 
{\longrightarrow} S_{\rm 3/2-parab}(\epsilon _{\kappa})~. 
\end{equation} 
Letaw plotted the numerical integral $S_{\rm hel}(\epsilon _{\rho})$, 
where $\epsilon _{\rho}=E/\rho$ for various values of $\kappa/\rho$.

\bigskip 
\noindent 
\underline{{\bf 4}. $\kappa=\tau _1$, $\tau _2=0$}, $\quad$ (if spatially projected: semicubical parabolas $y=\frac{\sqrt{2}}{3}\kappa x^{3/2}$) 

\bigskip

\noindent
The excitation spectrum is analytic, and 
since there are two equal curvature invariants one can use the 
dimensionless energy variable $\epsilon _{\kappa}$ 

\bigskip 
\begin{equation} 
S_{\rm 3/2-parab}(\epsilon _{\kappa})= \frac{\epsilon _{\kappa}^{2}}{8\pi ^2 \sqrt{3}} 
e^{-2\sqrt{3}\epsilon _{\kappa}}~. 
\end{equation} 
It is worth noting that $S_{\rm 3/2-parab}$, being a monomial times an exponential, 
is quite close to the Wien-type spectrum 
$S_{W}\propto\epsilon ^3e^{- {\rm const.}\epsilon}$.

\bigskip 
\noindent 
\underline{{\bf 5}.  $|\kappa|>|\tau _1|$, $\tau _2=0$, $\sigma ^2=\kappa ^2-\tau _1 ^2$} $\quad$ (if spatially projected: catenaries
$x=\kappa \cosh (y/\tau)$) 

\bigskip

\noindent
In general, the catenary spectrum cannot be found 
analytically. It is an intermediate case, which 
for $\tau/\sigma\rightarrow 0$ tends to $S_{\rm hyp}$, 
whereas for $\tau/\sigma\rightarrow\infty$ tends toward $S_{\rm 3/2-parab}$ 

\bigskip
\begin{equation} 
S_{\rm hyp}(\epsilon _{\kappa}) 
\stackrel{0\leftarrow \tau/\sigma}{\longleftarrow} 
S_{\rm catenary}(\epsilon _{\sigma})\stackrel{\tau/\sigma\rightarrow \infty} 
{\longrightarrow}S_{\rm 3/2-parab}(\epsilon _{\kappa})~. 
\end{equation} 

\bigskip 
\noindent 
\underline{{\bf 6}. $\tau _2 \neq 0$} $\quad$ (rotating worldlines uniformly accelerated normal to their plane of rotation)

\bigskip 

\noindent
The excitation spectrum is given in this case by a two-parameter set of curves. These trajectories are 
a superposition of the constant linearly accelerated motion and uniform 
circular motion. The corresponding vacuum spectra have not been calculated 
by Letaw, not even numerically.

\bigskip
\noindent
Thus, only the hyperbolic worldlines, having just one nonzero curvature 
invariant, allow for a Planckian excitation spetrum and lead to a strictly one-to-one 
mapping between the curvature invariant $\kappa$ and the `thermodynamic' 
temperature ($T_{U}=\kappa /2\pi$). The excitation spectrum due to  semicubical 
parabolas can be fitted by Wien-type spectra, the radiometric parameter 
corresponding to both curvature and torsion. 
The other stationary cases, being nonanalytical, lead to 
approximate determination of the curvature invariants, defining locally the 
classical worldline on which a relativistic quantum particle moves.

One can easily infer from these conclusions the reason why the Unruh effect became so prominent 
with regard to the other five types of stationary Frenet-Serret scalar spectra.

\bigskip
\noindent
I mention that Letaw introduced the terminology  {\em ultratorsional}, {\em paratorsional}, {\em infratorsional}, and {\em hypertorsional}
for the scalar stationary cases {\bf 3}-{\bf 6}. 


\bigskip
\noindent
\subsection{\bf The Electromagnetic Vacuum Noise} 

For the case of homogeneous electromagnetic field, Honig {\em et al} proved that the FS scalars remain constant along the worldline of a charged 
particle, whereas the FS vectors obey a Lorentz force equation of the form
\begin{equation} \label{L1}
\dot u ^{\mu}=\bar F^{\mu}_{\nu}u^{\nu}
\end{equation} 
with $F_{\mu \nu}=\lambda \bar F^{\mu}_{\nu}$, the electromagnetic field tensor, $\lambda =e/mc^2$, and  $\dot u ^{\mu}$ the four-velocity of the particle. 

A remarkable physical interpretation of the FS invariants in terms of the Lorentz invariants of the electromagnetic field was established by Honig {\em et al}
\begin{equation}\label{lorinv}
\kappa ^2-\tau _1^2-\tau _2^2  = \lambda^2(E^2-H^2)~,\qquad
\kappa \tau _2  =  -\lambda ^2(\vec{E}\cdot \vec{H})
\end{equation}

These beautiful results passed quite unnoticed until now. They could be used for a geometric transcription of homogeneous electromagnetism
and therefore for the geometric `calibration' of electromagnetic phenomena.

The FS formalism has not been used for the electromagnetic vacuum noise. Other approaches have been undertaken for this important case of which 
a rather complete one is due to Hacyan and Sarmiento that is briefly presented in the following.

\bigskip
{\em The Hacyan-Sarmiento approach}

\bigskip
\noindent
Starting with the expression for the electromagnetic energy-momentum 
tensor 
\begin{equation} 
\label{hs1} 
T_{\mu \nu}=\frac{1}{16\pi}\left( 
4F_{\mu \alpha}\;F_{\nu}^{\alpha}+ 
\eta _{\mu \nu}\;F_{\lambda \beta}\;F^{\lambda \beta} 
\right)~, 
\end{equation} 
Hacyan--Sarmiento define the electromagnetic two-point Wightman 
functions as follows 
\begin{equation} 
\label{hs2} 
D_{\mu \nu}^{+}(x,x')\equiv 
\frac{1}{4} 
\left( 
4F^\alpha{}_{(\mu}(x) \;F_{\nu )\alpha}(x')+ 
\eta _{\mu \nu}\; F_{\lambda \beta}(x)\; F^{\lambda \beta}(x')\right); 
\end{equation} 
\begin{equation} 
\label{hs3} 
D_{\mu \nu}^{-}(x,x')\equiv D_{\mu \nu}^{+}(x',x). 
\end{equation} 
This may be viewed as a variant of the ``point-splitting'' approach 
advocated by DeWitt.  Moreover, because of the properties 
\begin{equation} 
\label{hs4} 
\eta^{\mu \nu}\;D_{\mu \nu}^{\pm}=0, 
\qquad D_{\mu \nu}^{\pm}=D_{\nu \mu}^{\pm}, 
\qquad 
\partial _{\nu}D_{\mu}^{\pm \nu}=0, 
\end{equation} 
the electromagnetic Wightman functions can be expressed in terms of 
the scalar Wightman functions as follows 
\begin{equation} 
\label{hs5} 
D_{\mu \nu}^{\pm}(x,x')=c\; \partial_\mu \; \partial_\nu \;D^{\pm}(x,x'), 
\end{equation} 
where $c$ is in general a real constant depending on the case under 
study.  This shows that from the standpoint of their vacuum 
fluctuations the scalar and the electromagnetic fields are not so 
different.

Now introduce sum and difference variables 
\begin{equation} 
s = {t + t'\over2}; \qquad \sigma = {t-t'\over2}. 
\end{equation} 
Using the Fourier transforms of the Wightman functions 
\begin{equation} 
\label{ksy1} 
\tilde{D}^{\pm}(\omega , s)= 
\int _{-\infty}^{\infty} d\sigma \; 
e^{-i\omega \sigma}\; D^{\pm}(s,\sigma), 
\end{equation} 
where $\omega$ is the frequency of zero-point fields, the {\em 
particle number density of the vacuum seen by the moving detector} and 
the {\em spectral vacuum energy density per mode} are given by 
\begin{equation} \label{ksy2} 
n(\omega , s)=\frac{1}{(2\pi)^2 \omega} 
\left[ 
\tilde{D}^{+}(\omega , s )- 
\tilde{D}^{-}(\omega , s ) 
\right], 
\end{equation} 
\begin{equation} \label{ksy3} 
\frac{d e}{d\omega}=\frac{\omega ^2}{\pi} 
\left[ 
\tilde{D}^{+}(\omega , s )+ 
\tilde{D}^{-}(\omega , s ) 
\right]. 
\end{equation} 

\bigskip

{\em Circular worldline}

\bigskip
\noindent
The most important application of these results is to a uniformly 
rotating detector whose proper time is $s$ and angular speed is 
$\omega _{0}$ in motion along the circular world line 
\begin{equation} 
\label{hswl} 
x^{\alpha}(s)= 
(\gamma s, R_{0}\cos (\omega _{0}s), 
R_{0}\sin (\omega _{0}s), 0), 
\end{equation} 
where $R_{0}$ is the rotation radius in the inertial frame, 
$\gamma =(1-v^2)^{-1/2}$, and $v={\omega _0 R_0}/{\gamma}$. 
In this case there are two Killing vectors $k^{\alpha}=(1,0,0,0)$ and 
$ m^{\alpha}(s)=(0,-R_{0}\sin(\omega_{0}s), R_{0} 
\cos(\omega_{0}s,0)$. Expressing the Wightman functions in terms of 
these two Killing vectors, HS calculated the following physically 
observable spectral quantities (i.e., those obtained after subtracting 
the inertial zero-point field contributions):

\begin{itemize} 
\item {\em The spectral energy density} 
\begin{equation} \label{hsA} 
\frac{d e}{d\omega}= 
\frac{\gamma ^3}{2\pi ^3 R_{0}^{3}} \; 
\frac{\omega ^2+(\gamma v \omega _{0})^2}{\omega ^2} \; 
\frac{v^3w^2}{w^2+(2\gamma v)^2} \; 
h_{\gamma}(w), 
\end{equation} 
\item {\em The spectral flux density} 
\begin{equation} \label{hsB} 
\frac{d p}{d\omega}= 
\frac{\gamma ^3}{2\pi ^3 R_{0}^{3}}\; 
\frac{\omega ^2+(\gamma v \omega _{0})^2}{\omega ^2}\; 
4v^4 \; k_{\gamma}(w), 
\end{equation} 
\item {\em The spectral stress density} 
\begin{equation} \label{hsC} 
\frac{d s}{d\omega}= 
\frac{\gamma ^3}{2\pi ^3 R_{0}^{3}}\; 
\frac{\omega ^2+(\gamma v \omega _{0})^2}{\omega ^2}\; 
\frac{v^3w^2}{w^2+(2\gamma v)^2}\; 
j_{\gamma}(w). 
\end{equation} 
\end{itemize} 

\noindent
The ratio $(\omega ^2+(\gamma v \omega _{0})^2)/\omega ^2$ is a 
density-of-states factor introduced for convenience. 

\bigskip
\noindent
The Hacyan-Sarmiento variables are 
$$w=\frac{2\omega}{\omega _{0}} \qquad x=\frac{\sigma \omega _{0}}{2}$$ 

\noindent
$h_{\gamma}(w)$, $k_{\gamma}(w)$, and $j_{\gamma}(w)$ are the 
following cosine-Fourier transforms 
\begin{equation} 
\label{hsD} 
h_{\gamma}(w)\equiv   
\int _{0}^{\infty}\left(\frac{N_h(x,v)} 
{\gamma ^2[x^2-v^2\sin ^2 x]^3} 
-\frac{3}{x^4}+\frac{2\gamma ^2 v^2}{x^2}\right)\cos (wx)\; d x; 
\end{equation} 
\begin{equation} 
\label{hsE} 
k_{\gamma}(w)\equiv - 
\int _{0}^{\infty}\left(\frac{N_k(x,v)}{\gamma ^2[x^2-v^2\sin ^2 x]^3} 
-\frac{3}{x^4}-\frac{\gamma ^2}{6x^2}\right) 
\cos (wx)\; d x; 
\end{equation} 
\begin{equation} 
\label{hsF} 
j_{\gamma}(w)\equiv 
\int _{0}^{\infty}\left(\frac{1}{\gamma ^4[x^2-v^2\sin ^2 x]^2}- 
\frac{1}{x^4}+\frac{2\gamma ^2 v^2}{3x^2}\right)\cos (wx)\; d x. 
\end{equation} 
The numerators $N_h(x,v)$ and $N_k(x,v)$ are given by 
\begin{eqnarray} 
N_h(x,v)&=&(3+v^2)x^2+(v^2+3v^4)\sin ^2x-8v^2x \sin x; 
\\ 
N_k(x,v)&=&x^2+v^2\sin ^2x-(1+v^2)x \sin x. 
\end{eqnarray} 
%

\bigskip
{\em Ultra-relativistic limit}: $\gamma \gg 1$ 

\noindent
The functions
%
$ H_{\gamma}=\frac{v^3w^2}{w^2+(2\gamma v)^2}\; 
h_{\gamma}(w), 
\quad 
K_{\gamma}=4v^{4}\; k_{\gamma}(w), 
\quad 
J_{\gamma}=\frac{v^3w^2}{w^2+(2\gamma v)^2}\; j_{\gamma}(w)$, 
%
all have the following scaling property 
$X_{k\gamma}(kw)=k^3X_{\gamma}(w)$, 
where $k$ is an arbitrary constant, and $X=H,K,J$, respectively. This is the same 
scaling property as that of a Planckian distribution with a temperature proportional to $\gamma$.

\bigskip
\noindent

{\em  Nonrelativistic limit}: $\gamma \ll 1$ 

\noindent
A detailed discussion of the nonrelativistic limit for the scalar case has been provided 
by Kim, Soh, and Yee~\cite{KSY}, who used the parameters $v$ and 
$\omega _{0}$, and not acceleration and speed as used by Letaw and 
Pfautsch for the circular scalar case~\cite{LP}.  They obtained a 
series expansion in velocity 
\begin{equation} 
\label{ksy6} 
\frac{de}{d\omega}= 
\frac{\omega ^3}{\pi ^2} 
\left[ 
\frac{\omega _{0}}{\gamma\omega} 
\sum _{n=0}^{\infty} 
\frac{v^{2n}}{2n+1}\sum _{k=0}^{n}(-1)^{k} 
\; 
\frac{(n-k-\frac{\omega}{\gamma \omega _0})^{2n+1}}{k!\; (2n-k)!} 
\; 
\Theta\left(n-k-\frac{\omega}{\gamma \omega _0}\right)\right], 
\end{equation} 
where $\Theta$ is the usual Heavyside step function. Thus, to a specified 
power of the velocity many vacuum harmonics could contribute; making 
the energy density spectrum quasi-continuous.
For low frequencies the difference between the scalar and electromagnetic case is small.
Besides, one can consider only the first few terms in the series as an already good approximation.

\bigskip

\subsection{Non-stationary vacuum field noise} 

Non-stationary vacuum field radiation has a time-dependent spectral content requiring joint time and frequency information, i.e., one needs 
generalizations of the power spectrum concept. One can think of (i)  tomographical processing and/or (ii) wavelet transforms.  For 
instance, the recently proposed non-commutative tomography (NCT) transform $M(s;\mu,\nu)$~\cite{MVM}, if relativistically generalized, 
could be an attractive way of processing non-stationary signals.  In the definition 
of $M$, $s$ is just an arbitrary curve in the non-commutative time-frequency plane, 
while $\mu$ and $\nu$ are parameters characterizing the curve. The 
most simple examples are the axes $s=\mu t+\nu \omega$, where $\mu$ 
and $\nu$ are linear combination parameters.  The non-commutative 
tomography transform is related to the Wigner--Ville 
quasi-distribution $W(t,\omega)$ by an invertible transformation and has the following useful properties 
\begin{eqnarray} 
M(t;1,0)&=&|f(t)|^2, 
\\ 
M(\omega ;0,1)&=&|f(\omega)|^2, 
\end{eqnarray} 
where $f$ is the analytic signal which is simulated by $M$. 
Furthermore, employing $M$ leads to an enhanced detection of the 
presence of signals in noise which has a small signal-to-noise ratio. 
The latter property may be very useful in detecting vacuum field radiation noises, which are 
very small `signals' with respect to more common noise sources.

On the other hand, since in the quantum detector method, the vacuum 
autocorrelation functions are the essential physical quantities. In addition, 
according to various fluctuation-dissipation theorems these functions are 
related to the linear (equilibrium) response functions to an initial 
condition/vacuum. The fluctuation-dissipation approach has been 
developed and promoted by Sciama and collaborators~\cite{Sc}.  In 
principle, the generalization of the fluctuation-dissipation theorem 
for some classes of out of equilibrium relaxational systems, such as 
glasses, looks also promising for the case of non-stationary vacuum 
noise.  One can use a so-called two-time fluctuation-dissipation ratio 
$X(t,t')$ and write a modified fluctuation-dissipation relationship 
\cite{glass} 
\begin{equation} 
T_{\rm eff}(t,t') \; R(t,t')=X(t,t') \; 
\frac{\partial \tilde{C}(t,t')}{\partial t'}, 
\end{equation} 
where $R$ is the response function and $\tilde{C}$ the autocorrelation 
function.  The fluctuation-dissipation ratio is employed to perform 
the separation of scales. Moreover, $T_{\rm eff}$ are 
timescale-dependent quantities, making them promising for relativistic 
vacuum radiation, which corresponds naturally to out of equilibrium conditions.

\bigskip
\noindent
The necessity of a generalized nonequilibrium formalism, beyond Unruh effect, has been 
emphasized from a different standpoint by Hu and Johnson \cite{hj2000}.






\newpage
\section{\bf  DETECTION PROPOSALS}

\bigskip

A number of model experiments to detect `Unruh radiation' have been suggested in the last twenty years. The author has undertaken 
 the task of reviewing most of them \cite{rev1,rev2}. Here we present un update of our previous reviews.
Because the curvature thermodynamic temperature is given by $T_{\kappa}=\frac{\hbar}{2\pi c k}\,a$ this leads to $T_{\kappa}=4\cdot 10^{-23}\, a$
and one needs accelerations greater than $10^{20}g_{\oplus}$ to have `thermal' effects of only a few Kelvins. On the other hand, one should focus
below the Schwinger acceleration for copious spontaneous 
pair creation out of QED vacuum 
\begin{equation}\label{amax}
{a}_{\rm Schw} \approx m_e c^3/\hbar \approx 10^{29} \; m/s^2 
\approx 10^{28} \; g_{\oplus}~,
\end{equation}
There are indeed several physical settings in which accelerations can be achieved only a few orders 
below the Schwinger acceleration and forthcoming technological advances could test routinely those acceleration scales. The Unruh effect, if it exists,
can be revealed as a tiny signal in the background of by far more powerful effects.

\subsection{\bf  Unruh Effect in Storage Rings ($a\sim 10^{22}g_{\oplus}, \, T_{\kappa}=1200\, K$)}

\bigskip

\noindent
J.S. Bell and J.M. Leinaas imagined the first laboratory phenomenon connected to the
Unruh effect. During 1983-1987 they published a number of papers on the idea that the depolarising effects in
electron storage rings could be interpreted in terms of Unruh
effect \cite{bell,bell1,bel,mcd1}. 
However, the incomplete radiative polarization of the electrons in storage rings has been first predicted in early sixties in
the framework of QED by Sokolov and Ternov. Besides, the circular vacuum noise is not sufficiently ``universal" since it
always depends on both acceleration and velocity.
This appears as a `drawback' of the `storage ring electron thermometry', not to
mention the very intricate spin physics. Keeping these facts in mind,
we go on with further comments, following the nontechnical discussion of Leinaas \cite{lei}.

The circular acceleration in the LEP machine is $a_{LEP}=3\times 10^{22}g_{\oplus}$
corresponding to the Unruh temperature $T_{U}=1200$ $^{o}$K. It is a
simple matter to show that an ensemble of electrons in a uniform magnetic field at a nonzero temperature will have a polarization
expessed through the following hyperbolic tangent $ P_{U}=\tanh({\frac{\pi g}{2\beta}})$.
For the classical value of the gyromagnetic factor ($g=2$) and for
highly relativistic electrons ($\beta=1$),
$P_{U}=\tanh{\pi}=0.996$, beyond the limiting polarization of
Sokolov and Ternov $P_{ST, {\rm max}}=\frac{8\sqrt{3}}{15}=0.924$ \cite{sok}.

On the other hand, the function 
\begin{equation}\label{pu}
P_{U}(g)=\tanh({\frac{\pi g}{2\beta}})=\frac{1-e^{-\pi g}}{1+e^{-\pi g}}
\end{equation}
is very similar, when plotted, to the
function $ P_{DK}(g)$, which is a combination of exponential and polynomial
terms in the anomalous part of the gyromagnetic factor of the electron,
and it was obtained through QED calculations by Derbenev
and Kondratenko \cite{der}
\begin{equation}\label{DKf}
P_{eq}^{DK}=\frac{8}{5\sqrt{3}}
\frac{<\mid {\rho} \mid ^{-3}\hat {b} (\hat {n}- F_{DK})>}
{<\mid {\rho} \mid ^{-3}(1-2/9(\hat {n}\cdot \hat {v}) ^{2}+11/18
\mid F_{DK} \mid ^{2})>}~,      
\end{equation}
where $F_{DK}=\gamma\frac{\partial \hat {n}}{\partial \gamma}$ is the spin-orbit coupling function, which takes into account the
depolarizing effects of jumps between various trajectories differing from the reference closed orbit, $\rho$ is the bending radius, $\hat {b}$
a unit vector along the transverse magnetic field component. The brackets indicate an average over the ring circumference and over
the ensemble of particles in the beam. The unit vector $\hat {n}$ is the time-independent spin solution of the BMT equation, attached to each
particle trajectory. 

The difference between $P_U$ and $P_{DK}$ is merely a shift of the latter along the positive $g$-axis with about 1.2 units. As shown by Bell
and Leinaas, when the Thomas precession of the electron is included in the spin Hamiltonian, a shift of 2 units is obtained for $P_{U}(g)$.
This suggests a more careful treatment of spin effects arising when one is going from the lab system to the circulating coordinate frame.
A new spin Hamiltonian was introduced by Bell and Leinaas with a more
complicated structure of the field vector in the scalar product with the Pauli matrices. This complicated structure takes into
account the classical external fields, the quantum radiation field and the fluctuations around the classical path. Within this more
complete treatment, Bell and Leinaas were able to get, to linear order in the quantum fluctuations, a Thomas-like term and a third resonant
term. The latter is directly related to the vertical fluctuations in the electron orbit, which are responsible for the spin transitions
\begin{equation}\label{bl1}
P_{BL}=\frac{8}{5\sqrt{3}}\frac{1-\frac{f}{6}}{1-\frac{f}{18}+\frac{13}{360}f^2}
\end{equation}
where the resonance factor $f$  reads
\begin{equation}\label{fres}
f=\frac{2a_gQ^2}{Q_z^2-a_g^2\gamma ^2}~.
\end{equation}
$Q_z$ is the vertical betatron tune, $\gamma$ is the Lorentz factor and $a_g=(g-2)/2$ is the electron $g$ anomaly.
This induces an interesting variation of the beam polarization close to the resonance. As $ \gamma$ passes through it
from below, the polarization first falls from $92\%$ to $-17\%$, and then it increases
again to $99\%$ before stabilizing to $92\%$ . This is the only clear difference from the standard QED. Such resonances induced by the
vertical fluctuations of the orbit have been considered before in the literature within classical spin diffusion models and focused strictly on their depolarizing effect. 
Their nature is related to the fact that the Fourier spectrum of the energy jumps associated with the
quantum emission processes contains harmonics giving the usual resonance condition. As emphasized by Bell and Leinaas, a more direct
experimental demonstration of the circular Unruh noise would be the measurement of the vertical fluctuations. However, this will clearly
be a very difficult task since such fluctuations are among the smallest orbit perturbations. At the same time, the measurement
of the polarization variation close to the narrow resonance, in particular the detection of transient polarizations exceeding the Sokolov-Ternov limiting
one, will make us more confident in the claims of Bell and Leinaas. It is worth mentioning that the rapid passage through the resonance
does not change the polarization, while a slow passage reverses it but does not change the degree of polarization. Therefore only
an intermediate rate with respect to the quantum emission time scale of passing through the resonance will be appropriate for checking the transient
BL effects. 
Barber and Mane \cite{bar1} have shown that the DK and BL formalisms for the equilibrium degree of radiative electron
polarization are not so different as they might look. They also obtained an even more general formula for $P_{eq}$ than the DK and BL ones and
from their formula they estimated as negligible the BL increase near the resonance. It reduces to the following substitution of the resonance factor
\begin{equation}\label{barber1}
f\rightarrow f+\frac{2}{\gamma}~.
\end{equation}

The basic experimental data on spin depolarizing effects remain as yet
those measured at SPEAR at energies around 3.6 GeV in 1983. Away from the resonant
$\gamma's$ the maximum polarization of Sokolov and Ternov was confirmed \cite{john} but no
dedicated study of the polarization rates across resonances have been performed.




A paper of Cai, Lloyd and Papini \cite{reg} claims that the
Mashhoon
effect due to the spin-rotation coupling is stronger than the circular
Unruh effect (spin-acceleration coupling) at all accelerator energies
in the case of a perfect circular storage ring. However, the comparison
is not at all a straightforward one.

Bautista \cite{bau} solved the
Dirac equation in Rindler coordinates with a constant magnetic field
in the direction of acceleration and showed that the Bogolubov coefficients
of this problem do not mix up the spin components. Thus there is no spin
polarization due to the acceleration in this case.

In our opinion, the real importance of considering Unruh effect at storage
rings is related to clarifying radiometric features of the synchrotron
radiation \cite{ror}. There is a strong need to establish radiometric standards
in spectral ranges much beyond those of the cavity/blackbody standards,
and synchrotron radiation has already been considered experimentally
from this point of view \cite{ku}. Quantum field thermality is
intrinsically connected to the KMS condition. This is a well-known
skew periodicity in imaginary time of Green's functions expressing the
detailed
balance criterion in field theory. However, the task is to work out in more
definite terms the radiometric message of
the KMS quantum/stochastic processes \cite{kms}.

Recently, the spin-flip synchrotron radiation has been experimentally shown to be important
in the hard part of the spectrum in the axial channeling of electrons in the energy range
35-243 GeV incident on a W single crystal \cite{kirsebom}. This may revive the interest
in the BL interpretation, especially in the cleaner planar channeling case \cite{infou}. 

\bigskip


\subsection{\bf  Unruh Effect and Geonium Physics ($a\sim 10^{21}g_{\oplus}, \, T_{\kappa}=2.4\,K$)}  

\bigskip
\noindent
The very successful Geonium physics could help detecting the circular
thermal-like vacuum noise. The proposal belongs to J. Rogers \cite{rog}
being one of the most attractive.
The idea of Rogers is to place
a small superconducting Penning trap in a microwave cavity. A single
electron is constrained to move in a cyclotron orbit around the trap
axis by a uniform magnetic field (Rogers figure is B = 150 kGs).
The circular proper acceleration is $a= 6\times 10^{21}g_{\oplus}$
corresponding to T = 2.4 K. The velocity of the electron is maintained fixed
($\beta= 0.6$) by means of a circularly polarized wave at the electron
cyclotron frequency, compensating also for the irradiated power.
The static quadrupole electric field of the trap creates a quadratic
potential well along the trap axis in which the electron oscillates.
The axial frequency is 10.5 GHz (more than 150 times the typical
experimental situation \cite{bg}) for the device scale chosen by
Rogers. This
is the measured frequency since it is known \cite{bg} that the best way
of observing the electron motion from the outside world (Feynman's ``rest of
the Universe'') is through the measurement of the current due to the induced
charge on the cap electrodes of the trap, as a consequence of the axial
motion of the electron along the symmetry axis.
At 10.5 GHz the difference in energy densities between the circular
noise and the universal linear noise are negligible (see Fig. 2 in
Rogers' work). Actually, Rogers used the parametrization for the spectral
energy density of a massless scalar field as given by Kim, Soh
and Yee \cite{KSY} that he wrote in the form
\begin{equation} \label{rog1}
\frac{de}{d\omega}=\frac{\hbar}{\pi ^{2}c^{3}}\Big[\frac{\omega ^{3}}{2}+
\gamma \omega _{c}^{3}x^{2}\sum _{n=0}^{\infty}\frac{\beta ^{2n}}{2n+1}
+\sum _{k=0}^{n}(-1)^{k}\frac{n-k-x}{k!(2n-k)}\Theta [n-k-x]\Big]
\end{equation}
where $\gamma$ is the relativistic gamma factor,
$x=\omega/\gamma \omega _{c}$, and $\omega _{c}=eB/\gamma mc$ is the
cyclotron frequency. 
The power spectral density at the axial frequency is only
${\partial P}/{\partial f} = 0.47\cdot 10^{-22}$ W/Hz, and may be assumed
to be almost the same as the electromagnetic spectral energy density. This
power is resonantly transfered to the $TM_{010}$ mode of the
microwave cavity and a most sensible cryogenic GaAs field-effect
transistor amplifier should be used to have an acceptable
signal-to-noise ratio of S/N = 0.3. According to Rogers, the signal
can be distinguished from the amplifier noise in about 12 ms.

In conclusion, very stringent conditions are required in the model
experiment of Rogers. Top electronics and cryogenic techniques
are involved as well as the most sophisticated geonium methods. Taking
into account the high degree of precision attained by geonium
techniques, one may think of Rogers' proposal as one of the most feasible.
The critique of this proposal is similar to that in storage rings \cite{ros}
given that the circular Unruh effect is not universal, depending also on the
electron velocity. In addition, Levin, Peleg, and Peres \cite{lpp} studied the
Unruh effect for a massless scalar field enclosed in a two-dimensional circular
cavity concluding that the effects of finite cavity size on the
frequencies of normal modes of the cavity (Casimir effect) ignored by Bell
{\it et al} and by Rogers are in fact quite important.

\bigskip
{\em Cylindrical Penning Traps}

\noindent
A better experimental setting for detecting vacuum noises
by means of a trapped quantum detector (electron) may well be the cylindrical
Penning trap,
for which the trap itself is a  microwave cavity \cite{tg}. In this case
small slits incorporating choke flanges divide high-conductivity copper
cavity walls into the required electrode Penning configuration, including two
compensation electrodes. The driven axial resonance for this configuration
has already been observed with almost the same signal-to-noise ratio as
in hyperbolic Penning traps. By means of these cylindrical cavity traps,
a more direct coupling to the cavity modes may be achieved, especially in the
weak coupling regime. Here the cyclotron oscillator and the cavity mode
cannot form normal modes, and therefore other nonlinear effects are not coming
into play. The cylindrical $TM_{010}$ mode is essentially a zero-order Bessel
function in the radial direction and has no modes along the z axis. The price
to pay in the case of the cylindrical trap is a loss of control on the
quality of the electrostatic quadrupole potential.

\bigskip

\subsection{\bf  Unruh Effect and Nonadiabatic Casimir Effect  ($a\sim 10^{20}g_{\oplus},\, T\sim 1\, K$)}   

\bigskip
\noindent
An experimental equivalent of a {\em fast moving mirror} might be a
{\em plasma front} created when a gas is suddenly photoionized. This is
the proposal of Yablonovitch.\cite{yab} The argument is that
 the phase shift of the zero-point electromagnetic field
transmitted through a plasma window whose index of refraction
is falling with time (from 1 to 0) is the same as when reflected
from an accelerating mirror. Consider the case of hyperbolic motion.
Since the velocity is
\begin{equation}\label{yab1}
v= c\tanh(a\tau/c)
\end{equation}
where $\tau$ is the observer's proper time, the Doppler shift frequency
will be
\begin{equation}\label{yab2}
\omega_{D} = \omega_{0}\sqrt{\frac{1 - v/c}{1 +v/c}} =
\omega_{0}\exp({-a\tau/c})
\end{equation}
and consequently a plane wave of frequency $\omega_{0}$ turns into
a wave with a time -dependent frequency. Such waves are called chirped
waves in nonlinear optics and acoustics. Eq. (\ref{yab2}) represents an
{\em exponential chirping} valid also for black holes.
For an elementary discussion of
Doppler shift for accelerated motion see Ref.\cite{coc}.
It is worthwile to mention
that in the semiclassical treatment of black hole physics one
is usually dealing with chirped signals, since the WKB functions are
generally of
variable wavelength, and by meeting supplementary conditions on their
derivatives they are made to look as much as possible like fixed
linear combinations of plane waves.
On the other hand, in the case of wave packets
one is always working with the average frequency of the
wave packets (see the second paper of Jacobson \cite{jac} or the paper
of Frolov and Novikov on the dynamical origin of black hole entropy \cite{fn}).

The technique of producing plasma fronts/windows in a gas by laser breakdown,
and the associated frequency upshifting phenomena (there are also downshifts)
of the electromagnetic waves interacting with such windows,
are well settled since about twenty years.
Blue shifts of about $10\%$ have been usually observed
in the transmitted laser photon energy.

In his paper, Yablonovitch works out a very simple model of a {\em linear}
chirping due to a refractive index linearly decreasing with time,
$n(t)=n_{0}-\dot n t$, implying a Doppler shift of the form
$\omega \rightarrow \omega[1+\frac{\dot n}{n} t]\sim \omega[1+\frac{a}{c} t]$.
To have accelerations $a =10^{20}g_{\oplus}$ the laser pulses should be
less than 1 picosecond.
Even more promising may be the nonadiabatic photoionization
of a semiconductor crystal in which case the refractive index
can be reduced from 3.5 to 0 on the timescale of the optical
pulse. As discussed by Yablonovitch, the pump laser has to be tuned
just below the Urbach tail of a direct-gap semiconductor in order
to create weakly bound virtual electron-hole pairs. These pairs contribute a large
reactive component to the photocurrent since they are readily polarized. The
background is due to the bremsstrahlung emission produced by real electron-hole pairs, and to 
diminish it one needs a crystal with a big Urbach slope (the Urbach tail is an
exponential behavior of the absorption coefficient).

\bigskip
{\em Ambiguity of the interpretation}

\noindent
Yablonovitch remarked that
the experimental interpretation is highly ambigous. There are two types of reasoning:

\bigskip
(i) \underline{technologically-oriented interpretations}

\bigskip
\noindent
1. A single-cycle microwave squeezing 

\bigskip
\noindent
2. An inverse quadratic electro-optic effect with zero-point photons as input waves

\bigskip
(ii) \underline{concept-oriented interpretations}

\bigskip
\noindent
3. Nonadiabatic Casimir effect

\bigskip
\noindent
4. Unruh effect 

\bigskip
\noindent
Finally, one should notice the difference between the laboratory and
the black hole/hyperbolic chirping. The former is {\em linear}, whereas the latter is {\em exponential}.

\bigskip
\noindent
The `plasma window' of Yablonovitch was criticized in the important paper
by Dodonov,
Klimov, and Nikonov [DKN] \cite{dkn} on the grounds that we are not in the
case of exponentially small reflection coefficient as required to get
a Planck spectrum from vacuum fluctuations. At the general level,
one may argue that
nonstationary Casimir effects may produce some deformed Planck distributions,
and only in particular cases purely Planck distributions. As a matter of fact,
depending on the nonstationarity, one may obtain very peculiar photon spectra,
and this might be of great interest in applied physics.
DKN showed explicitly that an
exponential `plasma window', for which
presumably the modulation `depth' is the effective Unruh temperature, does
not produce a Planck spectrum. However, for a parametric function displaying
the symmetric Epstein profile one can get in the adiabatic limit a
`Wien's spectrum' with the effective temperature given by the logarithmic
derivative of the variable magnetic permeability with respect to time.
According to
DKN this corresponds to a `dielectric window' and not to a `plasma window'.
The experimental
realization of nonstationary Casimir effects are either resonators with
moving walls, as first discussed by Moore,\cite{mo} or resonators with
time-dependent refractive media as discussed by DKN. On the lines
of Yablonovitch, Hizhnyakov \cite{hiz} studied the sudden changes of the
refractive index caused by the excitations of a semiconductor near a
band-to-band transition in the infrared by a synchroneously pumped
subpicosecond dye laser, and also refered to the anology with Hawking
and Unruh effects. In 1994, C.K. Law \cite{law} combined the moving
walls of Moore with the dielectric medium with time-varying permittivity
in a one-dimensional electromagnetic resonant cavity. In this way, he obtained an effective
quadratic Hamiltonian, which is always required when we want to discuss
nonstationary `particle production' effects.


\bigskip

\subsection{\bf Unruh effect and sonoluminescence}

A few years ago, Eberlein elaborated more on Schwinger's interpretation of sonoluminescence in terms of zero point fluctuations and darely mentioned Unruh effect
\cite{eb1}.
She claimed that sonoluminescence could be interpreted as a generalized Unruh's effect that goes beyond the perfect mirrors restriction. In particular,
whenever {\em an interface between two dielectrics or a dielectric and the vacuum moves noninertially photons are created}.

Choosing as model profile for the time-dependence of the bubble radius about the collapse
\begin{equation}\label{sono1}
R^2(t)=R_{0}^{2}-(R_0^2-R_{\rm min}^2)\frac{1}{(t/\gamma)^2+1}~,
\end{equation}
Eberlein obtained for bubble radius $R$ much greater than the wavelengths of the light emitted the thermal-like spectral density
\begin{equation}\label{sono2}
S(\omega)=\frac{(n^2-1)^2}{64n^2}\frac{\hbar}{c^4\gamma}(R_0^2-R_{\rm min}^2)^2\omega ^3e^{-2\gamma \omega}~,
\end{equation}
using only zero-temperature quantum field calculations.calculations. She also calculated the total energy radiated during one acoustic cycle. She  obtained
$\sim 2\cdot 10^{-13}$ in a lapse of 1 femtosecond, claiming that this roughly corresponds to the observed number of photons. 
Her results were strongly contended by many comments. 
In particular, Milton \cite{milt96} used the Larmor formula $P=\frac{2}{3}\frac{(\ddot d)^2}{c^3}$ with the electric dipole moment $d\sim eR$
to estimate the classical Larmor energy radiated in the flash of  bubble collapse $E\sim \alpha \hbar c \frac{a^2}{c^3\tau ^3}$. This leads to only 10 eV for a flash of 1 femtosecond. To get $10^7$ eV,
the normal energy of the flash, one needs a time scale as short as $10^{-17}$ s, if Larmor radiation alone is invoked. Milton noted that models are required
in which the velocity of the radius is nonrelativistic while the acceleration becomes very large, and that this situation occurs during the formation of 
shocks

\bigskip

\subsection{\bf Unruh Effect and Squeezing in Fibers}

Grishchuk, Haus, and Bergman \cite{ghb92} discussed in a Phys. Rev. D paper a nonlinear Mach-Zhender configuration to generate radiation through the 
optical squeezing of zero-point fluctuations interacting with a moving index grating. 
The squeezed vacuum optical radiation has been generated experimentally by separating the pump light from the squeezed fluctuations in the MZ interferometric geometry.
The similarity with Unruh effect is mentioned in Section VIII of that paper.
There are two conditions to accomplish a strong optical squeezing in the laboratory: one needs to supress classical noise and phase match the vacuum wave with 
the exciting source. These two conditions are very well satisfied by working with fibers.

\bigskip

\subsection{\bf Unruh Effect and Channeling ($a\sim 10^{30}g_{\oplus},\, T_{\kappa} \sim 10^{11}\, K\, ?$)}

Relativistic particles can acquire extremely high transverse accelerations when they are channeled through crystals. 
Darbinian and collaborators \cite{dar} related this physical setting to Unruh radiation.

\bigskip
{\em First YERPHY Proposal}

\noindent
The idea is to measure the {\em Unruh radiation emitted in the Compton scattering 
of the channeled particles with the Planck spectrum of the inertial crystal vacuum}. The main argument
is that the crystallographic fields are acting with large transverse accelerations on the channeled particles.
The estimated transverse proper acceleration for positrons
channeled in the (110) plane of a diamond crystal is $a = 10^{25}\gamma$ cm/s$^{2}$, and
at a $\gamma = 10^{8}$ one could reach $10^{33}$ cm/s$^{2}$ = $ 10^{30}g_{\oplus}$.
Working first in the particle instantaneous rest frame, the Erevan
group derived the spectral angular distribution of the Unruh photons
in that frame. By Lorentz transformation to the lab system they got the
number of Unruh photons per unit length of crystal and averaged over
the channeling diameter. At about $\gamma = 10^{8}$ the Unruh intensity,
i.e., the intensity per unit pathlength of the Compton scattering on the
Planck vacuum spectrum
becomes comparable with the Bethe-Heitler bremsstrahlung
($dN_{\gamma}/dE\propto 1/E$, and mean polar emission
angle $\theta =1/\gamma$).

\bigskip
\noindent
Incidentally, there is a parallel with some experiments \cite{deh,bin,cer}
performed at LEP, where the scattering of the LEP beam from the
thermal photon background in the beam pipe has been measured (the
black body photons emitted by the walls of the pipe have a mean energy
of 0.07 eV). It is considered as a fortunate case that the effect is too small to affect the lifetime
of the stored beams.


\bigskip
{\em Second YERPHY Proposal}

\noindent
In another work of the armenian group \cite{dar1} the same type of calculations
was applied to get an estimate of the {\em Unruh radiation generated by TeV
electrons in a uniform magnetic field as well as in a laser field}.
The Unruh radiation becomes predominant over the synchrotron
radiation only when $\gamma = 10^{9}$ for $H = 5\cdot 10^{7}Gs$ and
consequently it is practically impossible to detect it at present linear colliders.
Supercolliders with bunch structure capable of producing magnetic
fields of the order $10^{9} G$ are required.
Pulsar magnetospheres are good candidates for considering such a Unruh
radiation.

\bigskip
\noindent
A circularly polarized laser field seems more promising since in this
case the Unruh radiation could be detectable at lower magnetic fields
and energies ($\gamma =10^{7}$). This is due to the fact that the proper
centripetal acceleration of the electron is
$a=2\omega\gamma\eta\sqrt{1+\eta ^{2}}$, where
$\omega $ is the frequency of the electromagnetic wave, and
$\eta =e\epsilon /m\omega$ ($\epsilon$ being the amplitude of the field).
\vspace*{0.2cm}

\bigskip

\bigskip

\subsection{\bf Unruh Radiation and Ultraintense Lasers ($a\sim 10^{25}g_{\oplus},\,T_{\kappa}= 1.2\, 10^{6}\,K$)}

\bigskip
\noindent
The most recent proposal is to get an Unruh signal in electron Petawatt-class laser interaction. It has been put forth by Chen and Tajima in 1999 \cite{ct99}.

Uniform acceleration through the usual quantum vacuum 
(Minkowski vacuum) of the electromagnetic field distorts the 
two-point function of the zero-point fluctuations (ZPF) in such a way 
that 
\begin{equation}\label{viss1}
\langle E_{i}(\sigma _{-}) E_{j}(\sigma _{+}) \rangle = 
{4\hbar\over\pi c^3} \; \delta_{ij} \; 
{(a/c)^4 \over \sinh^4(a \tau/2c)}. 
\hfill\qquad\qquad  
\end{equation}

The main point of Tajima and Chen is to introduce the so-called laser parameter (see below) in this formula.
In their approach  in the leading order the accelerated electron is assumed ``classical", with well-defined acceleration, velocity, and position.
This allows to introduce a Lorentz transformation so that the electron is described in its instantaneous proper frame. Also at this level the 
linearly accelerated electron will execute a classical Larmor radiation. As a response to the Larmor radiation, the electron reacts to the vacuum fluctuations with 
a nonrelativistic quivering motion in its proper frame that triggers additional radiation. The interaction Hamiltonian can be written as
\begin{equation}\label{ct99-2}
H_{int}=-\frac{e}{mc}\vec{p}\cdot \vec{A}=-e\vec{x}\cdot \vec{E}~.
\end{equation}
The probability of the emission of a photon with energy $\omega =\epsilon ^{'}-\epsilon$ is
\begin{eqnarray} 
N(\omega) & = & \frac{1}{\hbar ^2}\int d\sigma \int d\tau |\langle 1_{\vec{k}}, \epsilon ^{'}|H_{\rm int}|\epsilon, 0\rangle|^{2}\nonumber\\
& = &\frac{e^2}{\hbar ^2}\sum _{i,j}^{3}\int d\sigma \int d\tau e^{-i\omega \tau}\langle x_i(\sigma) x_{j}(\sigma)\rangle 
\,\langle E_{i}(\sigma _{-})
E_{j}(\sigma _{+})\rangle ~,
\end{eqnarray}
where $\sigma_{-}=\sigma -\tau/2$ and $\sigma_{+}=\sigma +\tau/2$ are combinations of $\sigma$ and $\tau$, the absolute and relative proper time, respectively. The $\tau$ dependence of the position operator has been extracted to the phase 
due to a unitary transformation. The last bracket is the autocorrelation function for the fluctuations of the electric field in the vacuum.

The laser is treated as a plane EM wave and since the interest is in a {\em quasilinear acceleration}, two identical, counterpropagating plane waves are considered in 
order to provide a standing wave. 

\begin{eqnarray}
E_x&=&E_0[\cos(\omega _0 t-k_0z)+\cos(\omega _0 t+k_0z)]~,\\
B_y&=&E_0[\cos(\omega _0 t-k_0z)-\cos(\omega _0t+k_0z)]~,
\end{eqnarray}
\noindent
To get the quasilinear acceleration and hopes for the Unruh signal, one should carefully tune the `laser' at the nodal points $k_0z=0,\pm2\pi ,...,$ where $B_y=0$ identically 
for all times and $E_x$ takes the maximum value. At $z=0$, Chen and Tajima find
\begin{equation} \label{ct99-6}
\gamma \beta _x=2a_0\sin (\omega _0 t) 
\end{equation}
with  
$$
\gamma =\sqrt{1+4a_0^2\sin ^2\omega _0 t}~, \quad {\rm and } \quad a_0=\frac{eE_0}{mc\omega _0}~,
$$
the relativistic $\gamma$ factor and the dimensionless laser strength parameter, respectively.

\bigskip

\noindent
The proper acceleration, which is related to that in the laboratory frame by $a=\gamma ^3 a_{\rm lab}$, is thus
\begin{equation} \label{ct99-7}
a=2ca_0\omega _0 \cos \omega _0 t~,
\end{equation}

\bigskip
{\em Autocorrelation function} 

\bigskip
\noindent
One has to look for the transformation between laboratory and proper spacetimes. As $d\tau =dt/\gamma$, in the limit $a\gg 1$ one finds
\begin{eqnarray}
\sin \omega _0 t  & = & \tanh (2a_0\omega _0 \tau)/\sqrt{1+4a_0^2{\rm sech} ^2(2a_0\omega _0 \tau)}\nonumber \\
\sin k_0x  & = & 2a_0 \cos \omega _0 t/\sqrt{1+4a_0^2}~.
\end{eqnarray}

\noindent
For periodic motion, it is sufficient to examine the cycle $-\pi/2 \leq \omega _0 t\leq \pi /2$ $\rightarrow$
$\pi/a_0\leq \omega _0 \tau \leq \pi/a_0$. Within the limit where $4a_0^2 {\rm sech}^2 (2a_0\omega _0 \tau) \gg 1$, the above equation reduces to
\begin{eqnarray}
\sin \omega _0 t & \approx &\frac{1}{2a_0}\sinh (2a_0 \omega _0 \tau)~,\nonumber\\
\cos k_0x  &\approx & \frac{1}{2a_0}\sinh (2a_0 \omega _0 \tau)~.
\end{eqnarray}

\noindent
In the limit of small arguments, these can be readily recognized as conformal transformations of the Rindler transformation for constant acceleration. 
Thus, the experiment corresponds to a {quasi-constant} acceleration and therefore one can use the known formula

\bigskip
\begin{equation}\label{ct99-10}
\langle E_{i}(\sigma -\tau/2)E_{j}(\sigma +\tau/2)\rangle =\delta _{ij}\frac{4\hbar}{\pi c^3}(2a_0\omega)^4{\rm csch}^4 (a_0\omega _0 \tau)~.
\end{equation}
From this, with some further mathematical tricks one arrives at the basic relationships

\bigskip
\begin{equation}\label{ct99-11}
\frac{dN(\omega)}{d\sigma}=\frac{1}{2\pi}\frac{e^2}{\hbar c^3}(2a_0\omega _0)^3\langle x^2\rangle\int _{-\infty}^{+\infty}ds\, e^{-is\omega/a_0\omega _0 }{\rm csch}^4 (s-i\epsilon)~,
\end{equation}

\bigskip
\begin{equation}\label{ct99-14}
\frac{dI_U}{d\sigma}=\int _{\omega _0}^{\infty}\hbar d\omega \frac{dN(\omega)}{d\sigma}\approx \frac{12}{\pi}\frac{r_e \hbar}{c}(a_0\omega _0)^3\log (a_0/\pi)~.
\end{equation}

\bigskip
\noindent
The above result applies to an accelerated electron located exactly at $z=0$. At the vicinity of this point, e.g., $k_0z=\epsilon \ll 1$, there is a nonvanishing magnetic field 
$|B_y|=\epsilon |E_x|$, which induces a $\beta _z \approx O(\epsilon)$ in addition to the dominant $\beta _x$. Nevertheless, the proper acceleration is affected only to the order $\epsilon ^2$
$$
a\approx 2c\omega _0 a_0[1-O(\epsilon ^2)]
$$

\bigskip
\noindent
For electrons farther away where the decrease of acceleration becomes more significant, both the Unruh and the background Larmor radiations will decrease much more rapidly due to their strong dependences on acceleration. 
Thus only the origin and its immediate surroundings are important within this proposal.

\bigskip

{\em Unruh Radiation versus Larmor Radiation}

\bigskip

\noindent
At the classical level, the same linear acceleration induces a Larmor radiation. 

\bigskip


\bigskip
\noindent
In addition the photon $\vec{k}$ space that they are interested in detecting is along the direction of acceleration where the Larmor radiation is the weakest.
Therefore the two radiations can be treated as independent processes without interference. 

\bigskip
\noindent
The total Larmor radiation power is
\begin{equation} \label{ct99-15}
 \frac{dI_L}{dt}=\frac{2}{3}\frac{e^2}{m^2c^3}\left(\frac{dp_{\mu}}{d\tau}\frac{dp^{\mu}}{d\tau}\right)=\frac{8}{3}r_emca_0^2\omega _0^2\cos ^2(\omega _0t)~.
\end{equation}
and the total energy radiated during each laser half-cycle is 
$$
\Delta I_L=\frac{4\pi}{3}r_emca_0^2\omega _0~.
$$
On the other hand, the Unruh radiation is significant over a reduced proper time period 
$$
\omega _0\Delta \sigma \geq O(1/a_0)
$$
Nevertheless, within this time the electron has become relativistic, with $\gamma \sim a_0$.
As a result, the total energy radiated in the lab frame, i.e., $\Delta I_{U}\sim (dI_U)/(d\sigma)\gamma \Delta \sigma$,
 is $\Delta I_U\sim \frac{12}{\pi}(r_e\hbar/c)a_0^3\omega _0^2\log(a_0/\pi)$. Thus the relative yield is
\begin{equation}\label{ct99-16}
\frac{\Delta I_U}{\Delta I_L}\sim\frac{9}{\pi ^2}\frac{\hbar \omega _0}{mc^2}a_0\log (a_0/\pi)~.
\end{equation}
Since $a_0\propto 1/\omega _0$, the relative yield is not sensitive to the laser frequency.

\bigskip
\noindent
1. {\em Petawatt-class laser}: $\omega _0\sim 2\cdot 10^{15}$ sec$^{-1}$ and $a_0\sim 100$ $\rightarrow
\frac{\Delta I_U}{\Delta I_L}\sim 3\cdot 10^{-4} 
$

\bigskip
\noindent
2. {\em Free-electron laser-driven coherent X-ray source}: $\hbar \omega _0\sim 10$ keV and $a_0\sim 10$ $\rightarrow
\frac{\Delta I_U}{\Delta I_L}\sim 1$

\bigskip
{\em Blind spot of Larmor radiation}

\bigskip
\noindent
The Unruh `thermal fluctuations' are considered isotropic in the electron's proper frame resulting in an isotropic induced radiation signal in the same frame.
But since at each half cycle the electron rapidly becomes relativistic, with $\gamma \sim a_0$, the Unruh radiation is boosted along the direction of 
polarization ($x$ axis) in the lab frame. Furthermore, as we have discussed above, the autocorrelation function, and therefore the Unruh signal, tend to 
diminish more rapidly than that from Larmor within the laser half cycle. This should induce a sharper time structure for the former.

Transforming the Unruh radiation power back to the lab frame with $\gamma \sim a_0$, the angular distribution in the small angle expansion becomes
\begin{equation}\label{ct99-19}
\frac{d^2I_U}{dtd\Omega}=\frac{4}{\pi ^2}\frac{r_{e}\hbar}{c}\frac{\omega _0^3a_0^3}{[1+a_0^2\theta ^2]^3}~.
\end{equation}
The Larmor radiation is polarized and its angular distribution in the small ($\theta , \phi$) polar-angle expansion is
\begin{equation}\label{ct99-20}
\frac{d^2I_L}{dtd\Omega}=\frac{8r_{e}mca_0^2\omega _0^2}{[1+a_0^2 \theta ^2]^3}\Big[1-\frac{4a_0^2\theta ^2(1-\phi ^2)}{[1+a_0^2\theta ^2]^2}\Big]~.
\end{equation}
The Larmor power is minimum at ($\theta , \phi)\,=\, (1/a_0 \pi , 0)$, where the angular distribution is zero. Consider a detector which covers an azimuthal 
angle $\Delta \phi =10^{-3}$ around the ``blind spot", and an opening polar angle, $\Delta \theta \ll 1/a_0$. Then the partial radiation power for the Unruh signal would dominate over that for the Larmor within this solid angle. A plot confirming this {\em blind spot} has been drawn by Rosu \cite{rpw}, 
whereas Visser \cite{viss01}
gave some interesting comments on the proposal of Chen and Tajima. I transcript here the following rethoric questions of Visser that he present as a sociology/linguistic issue:
``If you ultimately succeed in seeing this ZPF-induced modification to Larmor radiation, should you 
really call it the Unruh effect?  Or should you just call it basic quantum field theory? After all you are not directly measuring  the Unruh temperature itself."

\bigskip

{\em How to create ultralow electrons}

\bigskip
\noindent
1. {\em Low energy photoelectrons} near the surface of a solid material.

\bigskip
\noindent
2. {\em  Laser-trapped and cooled electrons}. Laser-electron interactions occur in vacuum and there should be minimal additional background effects.

\bigskip
\noindent
3. {\em  Low temperature plasma}. The main background effect is the bremsstrahlung. Even in this case the Unruh signal wins if the plasma density is low enough.
The cross section of bremsstrahlung for an unscreened hydrogen nucleus per unit photon energy is
$$
d\chi/d\hbar \omega \sim \frac{16}{3}\alpha r_e^2\ln (EE^{'}/mc^2\omega)~.
$$
The bremsstrahlung yield depends quadratically, whereas the Unruh signal linearly on the plasma density. 
The figure suggested by Chen \& Tajima is $n_p\leq 10^{18}$ cm$^{-3}$, below which the bremsstrahlung background can be minimized.

\bigskip
\subsection{\bf Unruh effect and damping in a linear focusing channel}

\bigskip
\noindent
K.T. McDonald \cite{McD00} applied recently the Unruh temperature formula for a rapid calculation of the damping in a linear focusing channel [LFC]
\footnote{ The LFC is a transport system that confines the motion of charged particles along straight central rays by means of a potential quadratic in the 
transverse spatial coordinates.}.
He used the same idea some 15 years ago to reproduce Sands' results on the limits of damping of the phase volume of 
beams in electron storage rings.


\newpage

\section{\bf  RELATED TOPICS}

\bigskip

\subsection{\bf  Unruh Effect and Anomalous Doppler Effect (ADE)}

\bigskip


\noindent
When studied with the detector method, the Unruh effect for a detector
with internal degrees of freedom is very close to the
anomalous Doppler effect (ADE). This is because in both cases the quantum detector
is radiating `photons' while passing onto the upper level and not on the
lower one \cite{frol}. 

\begin{figure}[htb]
\centerline{
\includegraphics[width=10cm,angle=0]{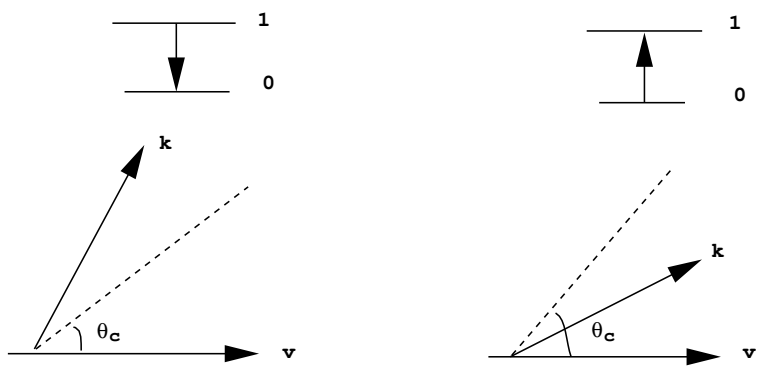}}
\caption{The normal and anomalous Doppler effects and the corresponding 
transitions.} \label{adef1}
\end{figure}

\bigskip
\noindent
It is worthwhile to note that the ADE-like concept
was used by Unruh and Wald \cite{wal} without referring to it
explicitly, when they considered the Unruh effect for
a uniformly accelerated quantum detector looked upon from the inertial
reference frame. Their main and well-known conclusion was that emission
in an inertial frame corresponds to absorption from the Unruh's `heat bath' in
the accelerated frame. Essentially one may say the following.

\bigskip
\noindent
(i) {\em For the observer placed in the noninertial frame the `photon'
is unobservable (it belongs to the left wedge in the Rindler case)}.

\bigskip
\noindent
(ii) {\em When the observer places himself in an inertial reference frame,
he is able to observe both the excited quantum detector (
 furnishing at the same time energy to it) and the `photons'. By
writing down the energy-momentum conservation law he will be inclined
to say that the `photons' are emitted precisely when the detector is
excited}.


\bigskip
{\em Review of ADE Literature}

\noindent
 A quantum derivation of the formula for the Doppler
effect in a medium has been given by Ginzburg and Frank already in 1947 \cite{47}
and more detailed discussion has been provided by Frank during 1970-1980 \cite{798}. 
See also the 1993 review paper of Ginzburg \cite{gi93}.

Neglecting recoil, absorption, and dispersion (a completely ideal case)
the elementary radiation events for a
two-level detector with the change of the detector proper energy
denoted by $\delta\epsilon$ are classified according to the photon
energy formula \cite{frol}
\begin{equation}\label{ade1}
\hbar\omega ={ -\frac{\delta\epsilon}{D\gamma}}
\end{equation}
where $\gamma$ is the relativistic velocity factor ($\gamma> 1$) and
D is the Doppler directivity factor
\begin{equation}\label{ade2}
D = 1 -( vn/c )\cos{\theta}
\end{equation}
The discussion of signs in Eq.~(\ref{ade1}) implies 3 cases as follows:

\bigskip

D$ >$ 0 for normal Doppler effect (NDE, $\delta\epsilon<0$)

D = 0 for Cherenkov effect (CE, $\delta\epsilon=0$, undetermined case)

D$ <$ 0 for anomalous Doppler effect (ADE, $\delta\epsilon>0$).

\bigskip

Consequently, for a quantum system endowed with internal degrees of
freedom the stationary population of levels is determined by the
probability of radiation in the ADE and NDE regions. The possibility
of doing population inversion by means of ADE has been tackled in the Russian
literature since long ago. A quantum system with many levels
propagating superluminally in a medium has been discussed for the
first time by Ginzburg and Fain in 1958 \cite{gf}.
The inverse population of levels by means of ADE or a combination of
ADE and acceleration may be enhanced whenever the ADE region is made
larger than the NDE region. This is possible, e.g., in a medium with
a big index of refraction.
Naryshkina \cite{nar} already found in 1962 that the radiation of
longitudinal waves in the ADE region is always greater than in the NDE
region. However, her work remained unnoticed until 1984, when
Nemtsov \cite{nem} wrote a short note on the advantage of using ADE
longitudinal waves to invert a quantum system propagating in an isotropic
plasma.
The same year, Nemtsov and Eidman \cite{nee} demonstrated inverse
population by ADE for the Landau levels of an electron beam propagating
in a medium to which a constant magnetic field is applied.
More recently, Kurian, Pirojenko and Frolov \cite{kpf} have shown that
in certain conditions (for certain range of the parameters), a detector
moving with constant superluminous velocity on a circular trajectory
inside a medium may be inverted too. Bolotovsky and Bykov \cite{bob}
have studied the space-time properties of ADE on the simple case of a
superluminous dipole propagating in uniform rectilinear motion in a
nondispersive medium. These authors are positive with the separate
observation of the ADE phenomenon for this case.

The radiation of a uniformly moving superluminal neutral polarizable
particle has been studied by Meyer \cite{mey}.
Frolov and Ginzburg \cite{frol} remarked that this case is an
analog of ADE due to zero-point fluctuations of electric polarizability.

Moreover, we can modify the index of refraction in the Doppler factor
in such a manner as to get the ADE conditions already at sublight
velocities. In this way a more direct link to the Unruh effect is
available, as has been shown also by Brevik and Kolbenstvedt\cite{bre}. 
These authors studied in detail the DeWitt detector moving through a dielectric nondispersive 
medium with constant velocity as well as with constant acceleration. In this way, first order perturbation theory 
formulas for transition probabilities and rates of emitted energy were exploited.

\bigskip
{\em Negative energy waves as induced ADE processes}

\noindent
Let us mention here that one way to look at negative energy waves in
plasma physics is to consider them as a manifestation of induced ADE
elementary events discussed in the book of Nezlin \cite{nez}.
As a matter of fact, a number of authors have already dealt with the problem
of amplification and generation of electromagnetic waves based on ADE in
the field of quantum electronics \cite{qe}. For details on the nonlinear
instabilities in plasmas related to the existence of linear negative
energy perturbations expressed in terms of specific creation and
annihilation operators, and also for a discussion of the complete
solution of the three-oscillator case with Cherry-like nonlinear coupling,
one should consult the Trieste series of lectures delivered by
Pfirsch \cite{pfi}.

\bigskip
{\em ADE elementary processes for channeled particles}

\noindent
Baryshevskii and Dubovskaya \cite{bdu} considered ADE processes for
channeled positrons and electrons.

\bigskip
{\em Negative energy waves in astrophysics/cosmology}

\noindent
Kandrup and O'Neill \cite{kon} investigated the hamiltonian structure of the Vlasov-Maxwell system in curved background
spacetime with ADM splitting into space plus time, showing the importance of negative energy modes for time-independent equilibria.

\bigskip


\subsection{\bf Unruh Effect and Hadron Physics}

The first application of the Unruh effect concept to hadron physics belongs to Barshay and Troost \cite{bt78} in 1978.
These authors identified the hadronic Hagedorn temperature with the Unruh temperature. In the same period,  
Hosoya \cite{hos} and also Horibe \cite{horib} applied moving mirror models to the thermal gluon production.
An estimation of the contribution of Unruh's effect to soft photon production by quarks is given in \cite{dim91}.
Dey {\em et al} \cite{dey93} related the Unruh temperature to the observed departure from the Gottfried sum rule for the difference of the 
proton and neutron structure functions in deep inelastic electron scattering. Also, Parentani and Potting \cite{pp89} commented on the 
relationship between the limiting Hagedorn temperature the maximal acceleration and the Hawking temperature. 

\subsection{\bf  Decay of Accelerated Protons}

\bigskip

\noindent

The stability of protons has been long used as a basic test for 
the standard model of elementary particles. 
The most recent high-precision experiment set $\tau _p > 1.6\cdot 10^{25}$ years \cite{caso}
higher than the present age of the universe. 
However, long ago Ginzburg and Syrovatskii \cite{gs} speculated on the possibilty of decay
in the case of noninertially moving protons.

\bigskip
\noindent
Recently, Vanzella and Matsas have calculated in the context of 
{\em standard QFT} (in inertial frames) the weak interaction decay rate for 
uniformly accelerated protons~\cite{VM}: 
\begin{equation} 
p^+  \to  n^0 + e^+_M  + \nu_M 
\label{pprocessinertial} 
\end{equation} 
and shown that in certain astrophysical situations the proton 
lifetime may be quite short. The energy necessary to render the process 
(\ref{pprocessinertial}) possible is supplied by the external 
accelerating agent. 
The tree-level formula for the 
proton proper lifetime in 1+1~spacetime 
dimensions ~\cite{MV} (valid under the no-recoil condition 
$a \ll m_p$) is
\begin{equation} 
{\tau}^{p  \to n}_{\rm Inert} (a) 
= 
\frac{2 \pi^{3/2} e^{\pi \frac{{\Delta m}}{a}}}{G_F^2  m_e } 
\left[ 
G_{1\;3}^{3\;0} 
\left( \frac{m_e^2}{a^2} \left| 
\begin{array}{l} 
\;\;\;1\\ 
-\frac{1}{2}\;,\;\frac{1}{2}+i \frac{\Delta m}{a}\;,\;\frac{1}{2}-i \frac{\Delta m}{a} 
\end{array} 
\right. 
\right) 
\right]^{-1} \; , 
\label{TIF} 
\end{equation} 
where 
$G_{p\; q}^{m n}$ is the Meijer function~\cite{GR}, 
$a$ is the proton proper acceleration, 
$ \Delta m \equiv m_n - m_{p }$ and $m_\nu = 0$ is assumed. (Here $m_p$, $m_n$, $m_e$ and $m_\nu$ 
are the rest masses of the proton, neutron, electron and neutrino, 
respectively.) 
The effective Fermi constant is $G_F=9.9 \times 10^{-13}$ 
is determined from phenomenology. 

Since a uniformly accelerated proton can be confined in a 
Rindler wedge which is a globally hyperbolic spacetime possessing 
a global timelike isometry, the associated uniformly accelerated 
observers (Rindler observers) must be able to analyze this 
phenomenon and reobtain the same (scalar) value for the proton lifetime 
(\ref{TIF}) 
obtained with standard QFT. Notwithstanding, because of energy 
conservation, Rindler observers would simply claim that protons are 
precluded from decaying into a neutron through 
\begin{equation} 
p^+  \to  n^0 + e^+_R  + \nu_R \; . 
\label{pprocessinertial'} 
\end{equation} 
If Unruh's thermal 
bath (Minkowski vacuum) is taken into account
new channels are opened: 

\newpage
\begin{equation} 
p^+ + e^-_R \to  n^0 + \nu_R 
\label{pprocessaccelerated1} 
\end{equation} 
\begin{equation} 
p^+ + \bar \nu_R  \to  n^0 + e^+_R 
\label{pprocessaccelerated2} 
\end{equation} 
\begin{equation} 
p^+  + \bar\nu_R + e^-_R \to  n^0  \; , 
\label{pprocessaccelerated3} 
\end{equation} 
i.e., from the point of view of the Rindler observers, the proton 
should be transformed 
into a neutron through the {\em absorption} of a Rindler electron and/or 
anti-neutrino from the surrounding thermal bath providing the 
necessary energy to allow the process to occur.
Any energy in excess can eventually be disposed by the emission of a neutrino 
or a positron (depending on the case).   
Performing an independent QFT calculation in the uniformly 
accelerated frame, Vanzella and Matsas
obtained the following proper lifetime for the proton, after combining 
(incoherently) processes 
(\ref{pprocessaccelerated1})-(\ref{pprocessaccelerated3}) 
in the presence of the vacuum thermal bath: 
\begin{equation} 
{\tau}^{p  \to n}_{\rm Rindler} 
= 
\frac{\pi^2 a e^{\pi \frac{\Delta m}{a}}}{G_F^2 m_e} 
\left[ 
\int_{-\infty}^{+\infty} d {\omega_R} 
\frac{ 
     K_{i \frac{{\omega_R}}{a} + \frac{1}{2}} ( m_e/a) 
     K_{i \frac{{\omega_R}}{a} - \frac{1}{2}} ( m_e/a)   
    } 
    {\cosh[\pi ({\omega_R} -{\Delta m})/a ] } 
\right]^{-1} 
\, . 
\label{TAF} 
\end{equation} 
Although Eqs.~(\ref{TIF}) and (\ref{TAF}) appear to be quite different, Vanzella and Matsas 
have shown numerically that they coincide up to the level
~\cite{VM2}: 
$$ 
\Delta^{-1} \int_\Delta 
[({\tau}^{p  \to n}_{\rm Rindler} 
- 
 {\tau}^{p  \to n}_{\rm Inert})/ 
 {\tau}^{p  \to n}_{\rm Rindler}]^2 dx
\sim 10^{-16}, 
$$ 
where $x \equiv \log _{10} (a/1 MeV)$.

\bigskip

\begin{flushleft} 
{\bf{\large Acknowledgements}} 
\end{flushleft}

\noindent
The author very much appreciates the invitation of Dr. Pisin Chen to QABP3 and the kindness of the hosts in Hiroshima.

\bigskip
\bigskip

\newpage

\end{document}